\documentclass[twocolumn,superscriptaddress,floatfix,showpacs]{revtex4-1}
\usepackage[dvips]{graphicx}
\usepackage{color}
\usepackage{amssymb,amsfonts,amsmath}

\newcommand{\bv}{{\bf {v}}}

\begin{document}
\title{Introduction to quantum turbulence}

\author{Carlo F. Barenghi}
\affiliation{Joint Quantum Centre Durham-Newcastle and School of 
Mathematics and Statistics, Newcastle University, i
Newcastle upon Tyne, NE1 7RU, United Kingdom}
\author{Ladislav Skrbek}
\affiliation{Faculty of Mathematics and Physics, 
Charles University, Ke Karlovu 3, 12116 Prague, Czech Republic}
\author{Katepalli R. Sreenivasan}
\affiliation{Department of Physics and Courant Institute of 
Mathematical Sciences, New York University, New York, NY 10012}

\begin{abstract}
The term quantum turbulence denotes the
turbulent motion of quantum fluids, systems such as
superfluid helium and atomic Bose-Einstein condensates which are
characterized by quantized vorticity, superfluidity and, at finite
temperatures,  two-fluid behavior.
This article introduces their basic properties,
describes types and regimes of turbulence which have been observed,
and highlights similarities and differences
between quantum turbulence and classical turbulence in ordinary fluids.
Our aim is also to link together the articles of this special issue,
and to provide a perspective of the future development of a subject which
contains aspects of fluid mechanics, atomic physics,
condensed matter and low temperature physics.
\end{abstract}


\maketitle
\section{Introduction}
Turbulence is a
spatially and temporally complex
state of fluid motion. Five centuries ago
Leonardo da Vinci noticed that water falling into a pond
creates eddies of motion.
Today turbulence
still provides physicists, applied mathematicians and engineers
with a continuing challenge.
Leonardo realized that the motion of water shapes the
landscape. Today's researchers appreciate that many physical processes,
from the generation of the Galactic magnetic
field to the efficiency of jet engines,
depend on turbulence.

The articles in this collection
are devoted to a special form of
turbulence known as quantum turbulence~\cite{PFrev,VinenNiemela,Nemir}, which appears in quantum fluids.
Quantum fluids differ from ordinary fluids
in three respects: (i) they exhibit two-fluid behavior
at nonzero temperature or in the presence of impurities,
(ii) they can flow freely, without the dissipative effect
of viscous forces, and (iii) their local rotation is constrained to discrete
vortex lines of known strength (unlike the eddies in ordinary fluids which are continuous
and can have arbitrary size, shape and strength).
Superfluidity and quantized vorticity are extraordinary
manifestations of quantum mechanics at macroscopic length scales.

Recent experiments have highlighted
quantitative connections as well as fundamental differences between
turbulence in quantum fluids and turbulence in ordinary
fluids (classical turbulence). The relation between
the two forms of turbulence is indeed a common theme in the articles
collected here. Since different scientific communities
(low temperature physics, condensed matter physics, fluid dynamics, atomic
physics) have contributed to progress in quantum turbulence\footnote{The term Quantum Turbulence was introduced into the literature
in 1986 by R.J. Donnelly in a symposium dedicated to the
memory of G. I. Taylor~\cite{DoSw}.},
the aim of this article is to introduce the main ideas
in a coherent way.

\section{ Quantum fluids}
In this series of articles we shall be concerned almost exclusively
with superfluid $^4$He, the B-phase of superfluid $^3$He, and,
to lesser extent, with ultra cold atomic gases.
These systems exist as fluids at temperatures
on the order of a Kelvin, milliKelvin and microKelvin,
respectively.\footnote{What determines
the need for a quantum mechanical description
is not the absolute value of temperature but whether it
is lower than a certain characteristic temperature of
the system (for example, the Fermi temperature);
instances of quantum fluids at high
temperature are exciton-polariton condensates ($\approx 300~\rm K$)
and neutron stars ($10^6$ to $10^9~\rm K$).} Their constituents
are either bosons (such as $^4$He atoms with zero spin)
or fermions (such as $^3$He atoms with spin 1/2).
This difference is fundamental: the former obey
Bose-Einstein statistics and the latter Fermi-Dirac quantum statistics.

Let us consider an ideal (non-interacting)
gas of bosons first. Under normal
conditions at room temperature, the
de Broglie wavelength $\lambda$ of each atom is much smaller
than the average separation $d$ between the atoms; if the temperature $T$ is
lowered, $\lambda$ increases, until, if $T$ is sufficiently small,
$\lambda$ becomes larger than $d$, and the quantum-mechanical
wave aspects become dominant. The resulting phase transition,
called the Bose-Einstein condensation~\cite{Annett}, is characterized by
a macroscopic number of bosons occupying the state of zero energy.
Although the possibility of Bose-Einstein condensation was raised
in 1924-1925, its direct experimental demonstration in dilute
ultra-cold atomic gases occurred only in 1995.
In order to achieve superfluidity (flow without friction) another
ingredient is necessary: the particles must interact with each other.

In fermionic systems, at temperatures much lower than a
characteristic Fermi temperature, particles occupy the
interior of the Fermi sphere in the momentum space,
with only relatively few particle-hole pairs, called excitations.
Attractive interaction between fermions leads to an instability
and the formation
of Cooper pairs which are bosons undergoing Bose condensation,
resulting either in superconductivity
in charged systems of electrons in crystal lattices, or in superfluidity
in systems consisting of neutral atoms.
The outcome is surprisingly similar to what happens for bosons: superfluidity arises from the formation of a coherent particle
field that can be described using the formalism
of the order parameter or a condensate macroscopic wave function.

Superfluidity of $^4$He was experimentally discovered by
Kapitza
and by Allen and Misener
in 1938, although
it is now believed that Kamerlingh Onnes must have had
superfluid helium in his apparatus when he first liquefied helium
in Leiden in 1908. He and other pioneers of low temperature physics
soon discovered that, below a critical temperature 
$T_{\lambda}\cong 2.17~\rm K$,
liquid helium displays unusual behavior. They therefore called it
\textbf{helium~I} and \textbf{helium~II},
respectively, above and below this temperature.
Still in 1938, London
linked the properties
of helium II to Bose-Einstein condensation.
Further progress in understanding  superfluidity of $^4$He was driven
by the work of Landau
based on different considerations.

The physical properties of normal liquid $^3$He
at milliKelvin temperatures are well described in the frame of the
Fermi liquid theory of Landau.
Note the striking difference in kinematic viscosities of $^4$He
(the lowest of all known fluids, two orders of magnitude less than
water's)
and of $^3$He near the superfluid transition
(comparable with that of air or olive oil).
Superfluidity of $^3$He was theoretically predicted by Pitaevskii
and experimentally discovered by Osheroff, Richardson and Lee in 1973.
Cooper pairs consisting of two $^3$He atoms (which themselves are fermions),
rotating about their center of mass, are bosons of total spin and orbital
numbers equal to one. This allows three different projections on
quantization axes in orbital and spin spaces and, as a consequence,
several different superfluid phases of $^3$He exist.
The classification of them and of the
types of quantized vorticity in $^3$He is beyond the scope of this article
(in particular because quantum turbulence has been studied only in the B phase).

Finally, recently developed experimental methods of laser and evaporative
cooling have opened up a
new road to ultra low-temperature physics: microKelvin clouds of dilute atoms were generated and
nanoKelvin temperatures achieved to explore quantum-degenerate gases,
providing additional working fluids to study quantum turbulence.

The two-fluid model, introduced in the context of
$^4$He first by Tizsa
and (based on different considerations) by Landau, is a
convenient level of description of quantum turbulence.
Below $T_{\lambda}$, $^4$He is described as a viscous normal fluid
(a gas of thermal excitations called phonons and rotons
that carry the entire entropy content)
coexisting with an inviscid superfluid
(related but not equal to the condensate fraction).
The density of helium II, essentially temperature independent,
can be decomposed into $\rho=\rho_n+\rho_s$, where
the normal fluid's and superfluid's densities,
$\rho_n$ and $\rho_s$, respectively, are strongly
temperature dependent: in the low temperature limit ($T \to 0$) helium
is entirely superfluid ($\rho_s/\rho \to 1$, $\rho_n/\rho \to 0$),
whereas, in the high temperature limit ($T\to T_{\lambda}$),
superfluidity vanishes ($\rho_s/\rho \to 0$,
$\rho_n/\rho \to 1$). At temperatures below 1~K (where
$\rho_n/\rho=0.07$),
in the absence of $^3$He impurities, $^4$He
can be considered more or less a pure superfluid. Similar considerations apply for superfluid $^3$He; its B phase
can be considered a pure superfluid below about 200~$\mu$K.

The normal fluid and the superfluid
support two independent velocity fields
$\bv_n$ and $\bv_s$, respectively, and
the superfluid component flows without viscous dissipation.
Based on the form of the dispersion relation, Landau predicted
that the superfluidity of $^4$He disappears at flow velocities exceeding
a critical value of about 60~m/s (due to the emission
of quasiparticles called rotons). In a more general sense, the Landau
criterion applies to any superfluid: on exceeding certain
critical velocity (which in fermionic superfluids is called
the Landau pair-braking velocity), it becomes energetically
favorable to generate quasiparticles, which means the onset of
dissipation.

What makes superfluid hydrodynamics particularly
interesting is that the circulation integral
$\oint_C \bv_s \cdot {\bf dr}$
is equal either to the quantum of circulation
$\kappa=h/m$ or to zero,
depending on whether or not the integration path $C$
encloses a quantized vortex line; here,
$h$ is Planck's constant and $m$
the mass of the relevant boson (one atom in $^4$He,
a Cooper pair in $^3$He-B). This quantization condition,
suggested by Onsager
and experimentally confirmed by Vinen, arises from the
existence and the single-valuedness
of a complex, macroscopic superfluid
wave function $\psi$, and
the usual quantum mechanical prescription that the velocity is proportional to
the gradient of the phase of $\psi$.  As a consequence,
the superflow is not only inviscid (like the ideal Euler fluid),
but also potential ($\nabla \times \bv_s=0$).
Vortex lines can be viewed as holes with circulation.
Moving around the vortex axis, the phase of $\psi$ changes by $2 \pi$
(multiple values of $\kappa$ are unstable in helium II), corresponding to
to a persistent azimuthal superfluid velocity of the form
$v_s=\kappa/(2 \pi r)$ where $r$ is the radial distance from the axis.
An isolated vortex line is thus a stable topological defect.
On its axis, real and imaginary parts of $\psi$ vanish;
the narrow region where the density drops from its value at infinity
to zero is proportional to the healing (or coherence)
length $\xi$, which depends on the strength of the interaction
between the bosons. In $^4$He,
$\xi \approx 10^{-10}~\rm m$;
in $^3$He-B, $\xi$ is about 100 times larger,
and in atomic condensates even larger
(1/100--1/10 of the system size).

Quantized vortex lines are nucleated
intrinsically or extrinsically (that is, from already existing vortex
lines, which, twisting under the influence of the superflow
and then reconnecting, generate new vortex loops).
Nucleation is opposed by a potential
barrier which, upon exceeding a critical velocity $v_c$,
can be overcome either thermally or by quantum tunneling.
In helium II (except close to $T_{\lambda}$) intrinsic nucleation requires
$v_c \approx 10~\rm m/s$, large enough to make it
unlikely unless induced by a fast-moving ion. Experimentally reported values of
$v_c \approx 10^{-2}~\rm m/s$ are associated with extrinsic nucleation
from remnant vortices pinned to the walls of the container.
In $^3$He-B both intrinsic and extrinsic nucleation are possible,
and (as in atomic condensates) $v_c \approx 10^{-3}~\rm m/s$.
Vortex lines can also form by the Kibble-Zurek
mechanism
when helium is cooled through the superfluid second order phase transition:
the phase of $\psi$, unable to adjust everywhere at the same
time, leaves vortex lines as defects.

The possibility of quantum turbulence
was first raised by Feynman \cite{Feynman};
soon afterward, Vinen showed that a turbulent vortex tangle can be
generated in the laboratory \cite{VinenOld} by applying a heat flux to helium II.
Two properties of vortex lines are important for
quantum turbulence. The first is the
mutual friction force~\cite{VinenOld,BDV} which couples
the superfluid and the normal fluid.
It arises from the scattering of thermal quasiparticles
(constituents of the normal fluid) by the velocity field of the
vortex lines. The second property comprises Kelvin waves, which are
helical displacements of the vortex core which
rotate with angular velocity $\omega \sim \kappa k^2$
where $k$ is the one-dimensional wavenumber (shorter waves rotate faster).
Kelvin waves arise from the
tension of the vortex lines (the kinetic energy
of a circulating superfluid about a unit length of line).
Their direct observation is reported in the article
by Fonda \emph{et al}.~\cite{FMQHL}.
At finite temperatures, Kelvin waves are damped by mutual friction, but
below $1~\rm K$ they propagate more freely and lead to acoustic
emission at large values of $k$. The
transfer of energy to such large $k$ by a Kelvin wave cascade
(analogous to the Kolmogorov cascade of classical turbulence)
explains the observed decay of turbulence at low temperatures,
as discussed in the articles by Barenghi, L'vov and Roche~\cite{BLR}
and by Walmsley, Zmeev, Pakpour and Golov~\cite{WZPG}; in the
weak amplitude regime, Kelvin waves can be studied using
wave turbulence theory, see the article by
Kolmakov, McClintock and Nazarenko~\cite{KMcN}.

The difference between the ideal fluid and the superfluid can be better
appreciated by noticing the link between
superfluidity and superconductivity, and the relation between
an ideal conductor and a superconductor; the latter behaves
as an ideal diamagnetic substance which, below certain critical conditions,
expels the externally
applied magnetic field from its interior (Meissner effect).
In superfluidity the corresponding feature is that the superflow
is always curlfree, or potential, independently on whether rotating
or quiescent sample were cooled through the superfluid transition.
Quantized vortices exist in superconductors and may reconnect
(the physical quantity which is quantized here is the magnetic flux,
in units $2\pi \hbar/(2e)$, where $2e$ is the
charge of two electrons constituting a Cooper pair).
The motion of vortex lines and flux
tubes, however, is not the same: if displaced, the former move (almost) along
the binormal and the latter (almost) along the normal direction \cite{Chapman},
which possibly explains the absence of quantum turbulence in superconductors.

\section{Experimental methods for quantum turbulence}
Some of the experimental methods used to probe
turbulence in ordinary viscous fluids have been used
for quantum turbulence. They include small Pitot tubes~\cite{MT}
to measure pressure head fluctuations (giving access to velocity
probability density distributions, structure functions and energy
spectra), and a plethora of small mechanical oscillators,
such as spheres, wires, nanowires, grids, quartz tuning forks
(for recent review, see~\cite{OscSV}) that can generate and detect
quantum turbulence.

Direct visualization is
invaluable in classical turbulence:
methods such as particle image
and particle tracking velocimetry, applied to scientific and industrial
problems, give
quantitative data and qualitative information such as flow patterns.
Although the application of these
methods to cryogenic flows is difficult for reasons which are both technical
(e.g., optical access to the experimental volume) and fundamental
(e.g., the presence of two velocity fields, interaction \cite{ChVanSciver}
of quantum vortices with particles---in most cases, micron sized frozen
flakes of solid hydrogen or deuterium), it has already led
to important results on the direct observation
of individual quantized vortices~\cite{Bewley},
their reconnections~\cite{Bewley2008},
Lagrangian velocity~\cite{Paoletti} and acceleration~\cite{MLM} statistics.
Another visualization technique~\cite{Guo}, based on fluorescence,
employs neutral He$^*_2$ triplet molecules as tracers.

An advantage of these conventional methods is that
they allow direct comparison between classical turbulence
above $T_{\lambda}$ and quantum turbulence below $T_{\lambda}$
but, in the latter case, care must be taken to understand whether
particles trace the
motion of the normal fluid, the superfluid, or the vortex lines.
The current status of the subject is described in the article by
Guo, La Mantia, Lathrop and Van Sciver~\cite{GMLS}.

Second sound attenuation is the most
powerful (and historically the oldest~\cite{VinenOld})
measurement tool in $^4$He, revealing the vortex
line density $L$ -- the total length of the quantized vortex line in a
unit volume ~\cite{Babuin}.
Since second sound is an anti-phase oscillation
of normal and superfluid components, this technique cannot be
used below $1~\rm K$ (since there is little normal fluid), or in $^3$He
at any temperature (second sound waves are
overdamped by the large viscosity of the normal fluid).

Helium ions have been successfully used to detect
quantum vortices in $^4$He---for example, to investigate
the decay of inhomogeneous quantum turbulence created by
ultrasonic transducers at about $1.5~\rm K$ \cite{MSS}.
An improved technique based on negative ions has been
recently introduced for measurements of decaying quantum turbulence
below $1$~K~\cite{Golov1,Golov2}.
Negative ions (electron bubbles) are injected by a sharp field-emission
tip and manipulated by an applied electric field. Bare
ions dominate for $T > 0.8$~K while, for $T < 0.7$~K, they
become self-trapped in the core of quantized vortex rings of
diameter about 1~$\mu$m (the rings are intrinsically
nucleated when the ions
are accelerated past $v_c$ by an imposed electric field).
Short pulses of ions or rings are sent across the experimental cell.
The relative reduction of the amplitude of pulses of ions or rings
detected at the collector
on the opposite side of the helium cell after their interaction
with quantum turbulence is converted to vortex line density.

In $^3$He-B, information on quantum turbulence
has been obtained using nuclear magnetic resonance~\cite{Nature}.
Another experimental technique in $^3$He-B is the
Andreev scattering of quasiparticles by the velocity field of
quantized vortices~\cite{BradleyDecay}, described in the article
by Fisher, Jackson, Sergeev and Tsepelin~\cite{FST}.

Finally, in atomic Bose-Einstein condensates,
vortices are created by stirring or shaking
the trap, phase imprinting, or moving a laser ``spoon" across the condensate;
images are taken after releasing the trap and
expanding the condensate, as explained by
White, Anderson and Bagnato~\cite{WAB}.

\section{Theoretical models of quantum turbulence}
Unlike classical turbulence, studied on the solid ground
of the Navier-Stokes equation, there is no single equation
governing the motion for quantum turbulence, but rather a
hierarchy of models at different length scales, each with its own
limitations. It is as if one is unable
to describe the trees and the forest in a unified way: we need one
(microscopic) model which
accounts for the close-up details of one or few trees,
a second (mesoscopic) model which, from further away,
does not resolve
the details of the trees but still distinguishes individual trees
as isolated sticks, and a third (macroscopic)
model which does not resolve trees at all but
recognizes where the forest is sparser or denser.
In this spirit we
note that helium turbulence is characterized by a wide separation of
length scales $\xi \ll \ell \ll D$, where
$\xi$ (already defined) is a measure of the vortex core,
$\ell$ is the average distance between vortex lines
(usually estimated as
$\ell \approx L^{-1/2}$), and $D$ is the size of the system;
typically
$\xi \approx 10^{-10}~\rm m$ in $^4$He ($10^{-8}~\rm m$ in $^3$He-B),
$\ell \approx 10^{-5}~\rm m$ and
$D \approx 10^{-2}~\rm m$. In atomic condensates, these scales
are not as widely separated: $\xi < \ell <D$.

The {\bf microscopic} model is the Gross-Pitaevski equation
for a Bose-Einstein condensate,
obtained (after suitable approximations) from the Hamiltonian of a Bose
gas undergoing two-bodies collisions. The Madelung transformation makes
the hydrodynamics interpretation of the wave function $\psi$
apparent, yielding the classical continuity equation and a modified Euler
equation. It differs from the classical Euler equation
because of the presence of the so-called quantum pressure, which,
differentiating a superflow from
a perfect Euler flow, is responsible for vortex reconnections
~\cite{Zuccher},
sound pulses at reconnection events~\cite{Leadbeater}
and for the nucleation of vortices near a boundary (e.g.
an ion~\cite{Pomeau}) or a
strong density variation (e.g. cavitation~\cite{MSS,Berloffbubble}).

The Gross-Pitaevskii equation
has been used to study turbulence
in atomic condensates \cite{Kobayashi2007,White2010}, but its
application to superfluid helium is only qualitative.
Its dispersion relation
does not exhibit the roton minimum
and is valid only near $T=0$.
For generalizations to finite temperature
we refer the reader to the article of Berloff, Brachet and Proukakis
~\cite{BeBraPr}. One approach worth
mentioning is the Zaremba-Nikuni-Griffin
formalism~\cite{Proukakis2008} which couples the Gross-Pitaevskii equation to
a Boltzmann equation for the thermal cloud of non-condensed atoms,
allowing atomic collisions within the thermal cloud and between
thermal cloud and
condensate: the outcome of this self-consistent interaction is
the emergence~\cite{Jackson2009} of
dissipative effects on vortex motion
(mutual friction).

A {\bf mesoscale} approach which shuns
effects at the
scale of $\xi$ is the vortex filament model of Schwarz~\cite{Schwarz}
which represents vortex lines as space curves
of infinitesimal thickness and circulation $\kappa$.
At $T=0$, a point on a vortex line
moves with the total superfluid velocity at that point---which is the
self-induced velocity calculated
using the Biot-Savart law, plus any externally imposed superflow.
For $T>0$, the motion results from the balance of
Magnus and friction forces.
Schwarz's insight was to recognize
that, in order to describe quantum turbulence,
his equation of motion must be supplemented with an algorithmic
procedure to reconnect vortex lines which
approach each other by a distance less than
the discretization distance along the lines (thus moving away from the realm of
Euler dynamics).

The vortex filament model is perhaps the most useful and flexible
numerical tool for quantum turbulence in $^4$He and $^3$He-B; the
state of the art is described by Baggaley and H\"anninen~\cite{BH}.
It is therefore important to appreciate its
limitations. The first is that (unlike the Gross-Pitaevskii equation)
it does not describe acoustic losses of energy
by rapidly rotating Kelvin waves at very low temperature.
The second limitation is that the computational cost of the
Biot-Savart law grows rapidly as $N^2$ (where $N$ is
the number of discretization points $N$ along the vortex lines).
To speed up his calculations,
Schwarz replaced the Biot-Savart law with its local induction approximation,
which neglects any vortex interaction and
requires an arbitrary mixing step to achieve
a statistically steady state of turbulence \cite{Adachi2010}---an approximation
which is thought as unsatisfactory by todays's standards.
This problem was recently solved~\cite{Baggaley-fluct,Baggaley-tree}
by adapting to vortex dynamics the $N\log{N}$ tree-algorithm
created for computational astrophysics \cite{BarnesHut}.
The third difficulty of Schwarz's model is that
(with notable exceptions \cite{Kivotides-ring,Kivotides-cloud})
the normal fluid velocity is imposed rather than computed
self-consistently by solving
the Navier-Stokes equation (suitably modified by a friction term):
again, the reason is the computational cost involved.

The problem of self-consistency
is solved, at a more {\bf macroscopic} level,
by the Hall - Vinen - Bekharevich - Khalatnikhov (HVBK) equations,
originally developed for rotating helium.
The HVBK equations describe the motion of fluid parcels containing
a large number of parallel vortex lines. Coarse-graining allows
treating superfluid
vorticity as a continuous classical field, generalizing the
original two-fluid equations of Landau.
The HVBK equations
successfully predict the oscillation of a rotating vortex
lattice, the Glaberson instability, the instability of helium
Couette flow and the transition to Taylor vortices
\cite{Couette,Henderson}, flows for which
the assumption is valid that the vortex lines are locally aligned within each fluid parcel.
Application of the
HVBK equations to turbulence is not justified
for randomly oriented vortex lines, as the
net superfluid vorticity in each
fluid parcel would be zero, yielding zero friction, despite the
nonzero vortex line density.
Modifications of the HVBK equations have been developed,
neglecting the vortex tension
and approximating the mutual friction~\cite{Leveque,Salort2011}.
Such models
probably underestimate friction dissipation but the coupled motion of
both fluids
is computed self-consistently.
Shell models of turbulence~\cite{Wacks,Lvov-SABRA} and
Leith models~\cite{LNS} are variants of the HVBK equations
trading spatial information for that in the k-space.

\section{Types and regimes of quantum turbulence}
It is useful to
classify the various types of turbulent flows which can be generated
in a quantum fluid, keeping in mind that more
classification schemes might well be possible.
To start with, the strong temperature
dependence of superfluid and normal fluid components and the relatively
high kinematic viscosity of $^3$He-B compared to $^4$He suggest the
following natural distinction:

\noindent
{\textbf{(i) Pure quantum (superfluid) turbulence
in low temperature $^4$He and $^3$He-B.}}
This is conceptually the simplest (but experimentally the most challenging) form of quantum turbulence: a single
turbulent superfluid (the normal fluid being absent or negligible).
This prototype of turbulence---a tangle of quantized vortex line---can be excited at small length scales by injecting ions or
vortex rings~\cite{Golov2}, or at larger length scales using
vibrating objects~\cite{OscSV,BradleyDecay}, or by suddenly halting the rotation
and destabilizing an existing vortex lattice~\cite{Golov1}.
Besides the residual friction potentially caused by thermal excitations, dissipation
of kinetic energy is possible due to acoustic emission from short and rapidly
rotating Kelvin waves \cite{JoeSound}
and from vortex reconnections~\cite{Zuccher}.
The length scale required for efficient acoustic emission
is much shorter than the typical curvature at the quantum
length scale $\ell \approx L^{-1/2}$, but can be achieved by
a Kelvin wave cascade---which is the energy transfer to increasingly smaller scales
arising from the nonlinear interaction
of Kelvin waves
\cite{Krstulovic-2012,Baggaley-Laurie-2013}; this mechanism is
discussed by Barenghi, Roche and L'vov ~\cite{BLR}). In $^3$He-B, the larger vortex core limits
the wavenumber range of the Kelvin wave cascade but
Caroli-Matricon bound states in the vortex core provide an additional
dissipation mechanism~\cite{Silaev}.

\noindent
{\textbf{(ii) Quantum turbulence with friction
in finite-temperature $^3$He-B.}}  The main feature of this form is that the highly viscous
normal fluid is effectively clamped to the walls.
The mutual friction force acts on all length
scales and affects the dynamics of quantized vortices,
damping the energy of Kelvin waves into the normal fluid.
The role of friction increases with rising temperature
to the point that, upon exceeding a critical temperature,
turbulence can be suppressed altogether.
Temperature thus plays an analogous role to that of the Reynolds number in classical turbulence
\cite{Nature}. This type of quantum turbulence is discussed
in the article by Eltsov, H\"anninen and Krusius \cite{EHK}.

\noindent
{\textbf{(iii) Two coupled turbulent fluids in $^4$He.}}
This type of turbulence is easily generated by stirring helium~II
by mechanical means (e.g., towed grids~\cite{SmithOregon,Stalp,SND} and propellers~\cite{MT}), by ultrasound~\cite{MSS}; or
by forcing it, using grids and flows past bluff bodies in wind tunnels~\cite{Salort2010}. Both superfluid and normal fluid component are turbulent.
Because of its double nature, this is the most general and challenging type
of quantum turbulence, generally richer than classical turbulence in conventional viscous fluids, presenting us with two coupled turbulent
systems, one in which the vorticity is continuous and the other with
quantized vortex lines. The mutual friction transfers energy from one fluid to the
other, so that it can act both as a source and a sink of energy for each fluid.
Moreover, by combining thermal and mechanical drives, special
types of turbulence can be generated in which the mean superfluid and
normal fluid velocities are not necessarily the same.
For example, thermal counterflow is induced
by applying a voltage to a resistor (heater) located
at the closed end of a channel that is open to a superfluid $^4$He
bath at the other end. The heat flux is carried away from the
heater by the normal fluid alone, and, by conservation
of mass, a superfluid current occurs in the opposite direction.
In this way a relative (counterflow)
velocity is created along the channel which is proportional to
the applied heat flux that is quickly accompanied by
a tangle of vortex lines \cite{VinenOld,Skrbek2003}.  The normal fluid is
likely to be laminar for small heat fluxes and probably turbulent for large heat fluxes.
Another special case is pure superflow~\cite{Babuin},
generated both mechanically
and thermally in a channel whose one or both ends are covered by
superleaks (walls with holes so tiny
that they are permeable only to the superfluid component).

A second possible classification of quantum turbulence
is based on the form of the energy spectrum $E(k)$ --
the distribution of kinetic energy over the
wavenumbers $k$ (inverse length scales).
Two limiting regimes have been tentatively identified:

\noindent (i) \textbf{Vinen} or \textbf{ultra-quantum}
turbulence. This is a random vortex tangle with a single dominant length
scale, $\ell$. It has long been argued \cite{VinenOld}
that steady counterflow turbulence at nonzero temperature in $^4$He
is in this regime, although the
energy spectrum has never been directly measured experimentally.
A recent numerical calculation \cite{Sherwin2012} of counterflow
turbulence driven by a uniform normal flow
has shown a broad energy spectrum around
$k \approx 1/\ell$, confirming this state.
At very low temperatures, ultra-quantum turbulence has been
produced in $^4$He by ion injection~\cite{Golov2}.
The main experimental evidence~\cite{Golov2,SreeniBook} is that,
if the vortex tangle is left to decay, the
vortex line density decreases as $L \sim t^{-1}$, in agreement with
a phenomenological model of Vinen~\cite{VinenOld} which indeed
assumes a random, homogeneous and isotropic vortex configuration.
The same $L \sim t^{-1}$ decay has been observed in numerical
simulations~\cite{Baggaley-ultra} which also confirmed that the
energy spectrum remains broadly concentrated near $k \approx 1/\ell$.

\noindent (ii) \textbf{Kolmogorov} or \textbf{semi-classical}
turbulence. This regime is quite similar to classical turbulence, as
the energy spectrum contains an inertial range and closely displays the celebrated K41 scaling
$E(k) \sim k^{-5/3}$ over $1/D \ll k \ll /\ell$
(hence most of the energy resides at the largest length scales).
Direct evidence of Kolmogorov scaling is provided by experiments
at high and intermediate temperatures~\cite{MT,Salort2010}
and numerical simulations~\cite{Sherwin2012} in which the vortex tangle
is driven by a turbulent normal fluid.
At very low temperatures the experimental evidence~\cite{Golov2}
is based on the observed decay $L \sim t^{-3/2}$
which, it has been argued~\cite{PFrev,VinenClassChar,SreeniBook},
is consistent with the Kolmogorov spectrum. At $T=0$,
numerical simulations based on both the vortex filament
model~\cite{Baggaley-ultra,Araki,Baggaley-structures} and the
Gross-Pitaevskii equation~\cite{Nore,Kobayashi2005} have produced spectra
consistent with the $k^{-5/3}$ scaling. Further
numerical studies have revealed that the Kolmogorov energy spectrum is
associated with the presence of metastable bundles of
polarized quantized
vortices~\cite{Laurie2012,Baggaley-structures,Sherwin2012}.
This opens the possibility of
stretching such bundles (stretching of individual quantum
vortices is not possible because of the quantization condition).
The polarization of (part of) the vortex tangle is
also discussed in the article of Vinen and Skrbek~\cite{VS}
on turbulence generated by oscillating objects; its importance lies on
the fact that in classical turbulence vortex stretching
is thought to be responsible for the dissipationless transfer of energy
from large to small scales (energy cascade).

An important issue
is the normal fluid's profile in various types of channel and pipe flows of helium II. For example, in numerical simulations
of counterflow turbulence driven by uniform normal fluid,
the energy spectrum broadly peaks at the
mesoscales $k \approx 1/\ell$~\cite{Sherwin2012} (as in ultra-quantum
turbulence), but the experimentally
observed decay is $L\sim t^{-3/2}$~\cite{Skrbek2003} (typical of
of quasi-classical turbulence).
Thus either large scale structures already exist in steady-state
counterflow or are generated by halting the normal fluid.

A third classification is suggested by the relative magnitude of
$\xi$, $\ell$ and $D$.  \textbf{Quantum turbulence
in atomic Bose-Einstein condensates} lacks the wide separation of
scales typical of liquid helium: the size of typical condensates
is only 10 to 100 times the healing length.  Will
the known scaling laws of turbulence manifest themselves as the size of the
condensate increases? The close distance between
vortices suggests that reconnections play a more important
role in the kinetic energy dissipation than in helium; moreover,
vortex energy can be transformed into surface oscillations
of the condensate.
Atomic condensates can be used to explore the cross-over from
two-dimensional to three-dimensional turbulence, offer greater
flexibility than helium as physical parameters can be engineered and
vortices can be individually manipulated,
and, thanks to the weak interactions, are a testing ground for the theory.
Two-components \cite{Takeuchi}
and spinor condensates \cite{TsuBose} are rich new systems for turbulence.
Dipolar condensates \cite{Mulkerin}
may open the possibility of turbulence with
unusual vortex interaction.
These opportunities are discussed in the article by White, Anderson
and Bagnato~\cite{WAB}.

\section{Outlook}
Quantum turbulence is a relatively young field of research
compared with conventional turbulence in viscous fluids, which
has slowly but steadily progressed over several centuries.
The early works on quantum turbulence~\cite{VinenOld} were
mainly concerned with counterflow as a
problem of heat transfer unique of liquid helium II.
It was only after the seminal experiments of Donnelly, Tabeling
and collaborators \cite{SmithOregon,Stalp,SND,MT} that the attention
shifted to concepts such as energy spectrum and vorticity
decay, which are typical of the fluid dynamics literature.
These and other experiments showed that, over length scales much
larger than the mean intervortex distance $\ell$,
quantum turbulence mimics~\cite{VinenClassChar,SreeniBook}
the properties of classical turbulence, hinting (in the spirit of
Bohr's old quantum theory) that many quanta of
circulation yield classical behaviour.
This result, together with the very low kinematic
viscosity of $^4$He, suggests that quantum turbulence can be used
to study classical problems such as the
temporal decay of homogeneous, isotropic turbulence,
or the
long-standing puzzle of the Loitsianskii invariant.
In general, highly turbulent flows are needed to tackle these problems.
It is not difficult to generate such flows with
liquid $^4$He, and
CERN's huge capacity liquefiers are already considered for the purpose
within the European project EuHit, in the frame of 7th EU
initiative. A more challenging task is the development of
miniature special probes capable of probing quantum turbulence,
resolving all scales including quantum scales smaller than $\ell$.

On the other hand, the existence of the ultra-quantum regime is a warning
signal that not all quantum turbulent flows are related to classical flows.
The various types and regimes of quantum turbulence
which we have identified
provide a rich range of problems
which we should solve using hydrodynamics models (the spirit is similar to how
the known planetary atmospheres are explained by the same physical
principles under different parameters).

Problems which seem particularly challenging involve either
two turbulent cascades taking place in the same fluid in different
regions of k-space (the Kolmogorov cascade and the Kelvin waves
cascade), or two active
turbulent superfluids affecting each other (e.g. two-component cold
gases \cite{Takeuchi} and, when experimentally realized \cite{Tuoriniemi}, 
$^3$He-$^4$He mixtures
with both $^3$He and $^4$He superfluid). For complexity and difficulty, the closest analogue in classical
physics is perhaps the problem of coupled turbulent velocity and
magnetic fields in astrophysical magneto-hydrodynamics.

The temperature of the cosmic microwave background radiation
is about $2.7~\rm K$ and the coldest place found in the Universe
is the Boomerang Nebula ($\approx 1~\rm K$), 5000 light-years away from
Earth in the constellation of Centaurus.
Thus further experimental
studies of quantum turbulence, probing physical
conditions not known to Nature at temperatures many orders
of magnitude lower, may uncover phenomena not yet known
to physics.


\begin{acknowledgments}
We thank W.F. Vinen for stimulating discussions and constructive criticism.
LS acknowledges support of the Czech Science Foundation under
GA\v{C}R 203/14/02005S and by the European Commission under the 7th Framework Programme EuHIT. CFB is grateful to the EPSRC for grant EP/I019413/1.
\end{acknowledgments}

\end{document}